\documentclass{elsart5p}

\usepackage{amssymb}

\usepackage{lineno}

\usepackage{epsfig}
\begin{document}

\begin{frontmatter}



\title{AURA - A radio frequency extension to IceCube}

\author[uw]{H.~Landsman} for the IceCube Collaboration,
 \author[uh]{ L.~Ruckman},
\author[uh]{ G.S.~Varner}
\address[uw]{Dept. of Physics, Univ. of Wisconsin, Madison, WI 53703, USA}
\address[uh]{Dept. of Physics and Astronomy, Univ. of Hawaii, Manoa, HI 96822, USA}

\begin{abstract}
The excellent radio frequency transparency of cold polar ice, combined with the coherent Cherenkov emission produced by neutrino-induced showers when viewed at wavelengths longer than a few centimeters, has spurred considerable interest in a large-scale radio-wave neutrino detector array. The AURA (Askaryan Under-ice Radio Array) experimental effort, within the IceCube collaboration, seeks to take advantage of the opportunity presented by IceCube \cite{icecube}\cite{icecube2} drilling through 2010 to establish the radio frequency technology needed to achieve $100-1000$ $\mathrm{ km^3} $ effective volumes. 
In the 2006-2007 Austral summer 3 deep in-ice radio frequency (RF) clusters were deployed at depths of $\sim1300$ m and $\sim300$ m on top of the IceCube strings. Additional 3 clusters will be deployed in the Austral summer of 2008-2009. Verification and calibration results from the current deployed clusters are presented, and the detector design and performances are discussed. 
Augmentation of IceCube with large-scale $(1000$ $\mathrm{ km^3sr})$ radio and acoustic arrays would extend the physics reach of IceCube into the EeV-ZeV regime and offer substantial technological redundancy.\end{abstract}

\begin{keyword}
  Neutrino astronomy \sep Radio Frequency \sep RF \sep Ice \sep South Pole \sep GZK Neutrinos \sep Neutrinos detection 
\PACS 
\end{keyword}
\end{frontmatter}

\section{Introduction}
Astrophysical high energy neutrinos might carry valuable information about their sources, possibly point sources like GRBs, AGNs, or SGRs, or high energy cosmic rays (through the GZK process). They might also offer the chance to investigate particle physics in an energy range unreachable by Earthbound accelerators


Recent measurements from cosmic ray detectors indicate a change in the CR spectrum as predicted by the GZK process \cite{Auger}.
The expected flux for GZK neutrinos is so low that IceCube is expected to measure no more than a few events a year \cite{ESS} and the suggested heavier composition of the cosmic rays will further lower this flux.
 Measuring GZK neutrinos will help explain the generation mechanism of the CRs, their flux and composition.


Current neutrino detectors like IceCube, AMANDA, and Antares incorporate hundreds of photo-multiplier tubes, sensitive to photons in the optical wavelength range. They are designed to detect neutrinos with energies between $10^{2}$ GeV -$10^{10}$ GeV . In order to survey the extreme high energy regime of above $10^{10}$ GeV, larger detectors will be needed.


In 1963, G.A.Askaryan \cite{askaryan} suggested that cascades generated by high energy charged particles moving through a dielectric at relativistic speeds, build up an excess of negative charge, giving rise to strong coherent radio and microwave Cherenkov emission. Coherence obtains when the wavelength of the emitted radiation is longer than the transverse dimensions of the cascade. It is expected that neutrinos with energy of $\sim10^{18}$ eV or more will produce cascades with transverse dimensions of order $\sim0.1$ m, thus emitting coherent radio frequency (RF) radiation. This effect was demonstrated in an accelerator measurement where coherent linearly polarized RF radiation was measured from the interaction of a beam dumped into RF transparent matter (sand, salt and ice)\cite{slac}.

As the limits on possible fluxes of extraterrestrial neutrinos measured by current experiments become more stringent, it becomes evident that even larger detectors are needed for gathering enough statistics on human time scales.

The cost of hardware, deployment and drilling in the ice limit the size of IceCube and alternatives to optical detection are needed to cover larger volumes.  The simpler installation of radio detectors, the long attenuation length of RF in shallow ice and 
the sensitivity to extreme high energy (EHE) events, together with lower cost as compared to optical detectors, makes the RF region interesting for EHE neutrino detection.

The concept of a GZK radio frequency detector, deployed at shallow depths or as a surface array had been suggested more than 20 years ago \cite{array}. Several experiments are already using the Askaryan effect for neutrino detection in Antarctica: The RICE\cite{rice}  array was deployed with the AMANDA neutrino telescope near the South Pole at depths of 100-300 m. The array consists of 20 dipole antennas covering a volume of $200\times200\times200$ $\mathrm{ m^3}$, and is sensitive between 200 and 500 MHz. RICE not only established the feasibility of a radio array and measured the RF properties of South Pole ice, but also published limits on neutrino fluxes between $10^{16}$ eV and $10^{18}$ eV  and on exotic models. The ANITA\cite{anita} experiment, balloon borne at 40 km, observes the Antarctic ice searching for RF emission. The high altitude allows ANITA to cover a large volume (1.5 million $\mathrm{ km^3}$), but the short flight time limits its exposure. 

AURA builds on the experience of the RICE experiment with its electronic design based on the RF specific electronic applications developed by ANITA \cite{labrador}. We use the communication and time calibration systems developed for IceCube and rely on the experience within the IceCube collaboration to develop hardware and 
procedures for building and deploying highly sensitive equipment in the extreme environment of the South Pole.


\section{South Pole ice}
\begin{figure}
\begin{center}
\noindent
\includegraphics*[width=10cm]{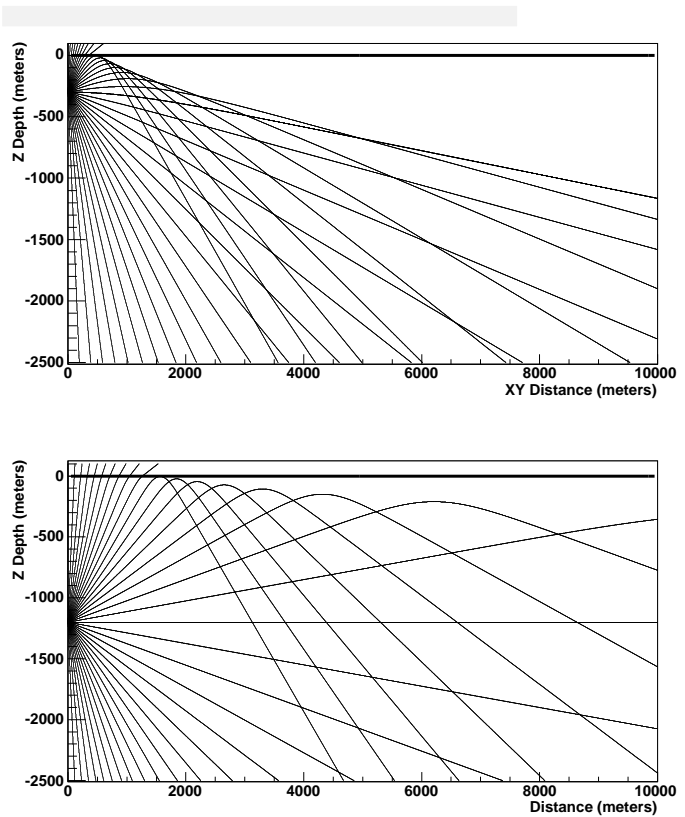}
\end{center}
\caption{Ray trace modeled for detectors deployed at -300 m (top) and -1200 m (bottom). Only waves propagating at certain angles will emerge to the surface.}
\label{fig:n}
\end{figure}

The RF ice properties determine the feasibility and design of a future GZK detector. Specifically the attenuation length will influence the spacing between channels, and the index of refraction will determine the reconstruction capabilities and simulation quality. The latest index of refraction measurement is reported in \cite{ice-n} using the RICE array down to 150 m and ice cores down to 240 m.  
The index of refraction changes rapidly in the soft ice layers on top of the glacier (firn) and decreases the angular acceptance of shallow deployed detectors by causing total reflection of rays propagating between the layers. This is illustrated in Fig. \ref{fig:n}.

On the other hand, the attenuation of RF decreases with temperature, making colder 
ice more RF transparent. Thus shallow deployment in colder ice, is more favorable for RF 
detection. This measurement is summarized in \cite{ice-atten} where a surface transmitter was used to send signals down into the ice. A receiver recorded the signals after bouncing back from the bedrock. This measurement was performed at frequencies of 200-700 MHz and provided an average RF attenuation for the round-trip down and back.
The attenuation as a function of depth is then calculated using temperature models. The IceCube detector provides an accurate temperature measurement down to 2.5 km. The average attenuation length is 1.5 km, with longer lengths for lower frequencies, and for shallower (colder) ice.

In the coming season, using a powerful in-ice transmitter unit we will try to perform point-to-point direct measurement of attenuation length.

\section{The detector design}
\subsection{AURA I : 2006-2007 Design}
\begin{figure}
\begin{center}
\noindent
\includegraphics*[width=5cm]{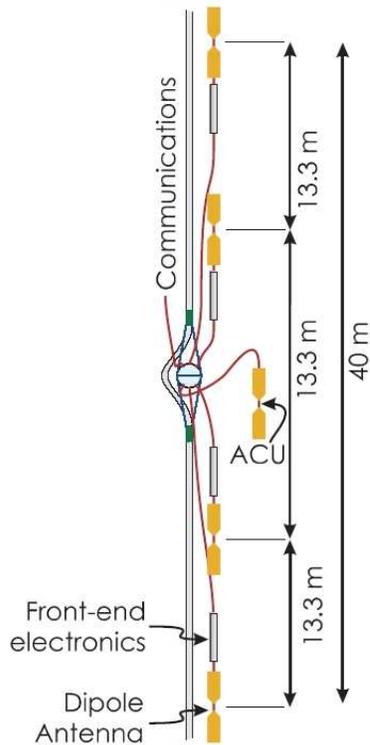}
\end{center}
\caption{The radio cluster, made of a  DRM (Digital Radio Module), and 5 antennas (4 receivers and a transmitter).}
\label{fig:cluster-r}
\end{figure}

Each AURA unit, called a ``cluster'', consists of 4 receiver antennas, equally spaced along 40 m, a transmitting antenna for calibration and a sphere containing the electronics, called a DRM - Digital Radio Module.  Six cables are connected to the DRM: One for power and communication going to the surface and to a computer  handling data acquisition and control; an additional cable holds the Array Calibration Unit (ACU); and four cables connected to receiver antennas. A schematic of the AURA cluster is shown in Fig. \ref{fig:cluster-r}. A set of front-end electronics, including a chain of amplifiers and filters is 
mounted before each of the four receivers, between the antennas and the DRM.  

\begin{figure}
\begin{center}
\noindent
\includegraphics*[width=10cm]{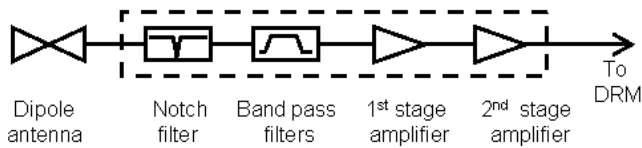}
\end{center}
\caption{The front end electronics inside a metal tube consist of a set of filters and amplifiers.}
\label{fig:frontend}
\end{figure}

The DRM, within a $36$ cm diameter glass sphere, contains the triggering, digitization and communication electronics as well as a power converter. It holds the TRACR board (Trigger Reduction And Communication for RICE) that controls the calibration signal and the high triggering level, the SHORT board (SURF High Occupancy RF Trigger) that provides frequency banding of the trigger source,  the ROBUST card (Read Out Board UHF Sampling and Trigger) that provides band trigger development, high speed digitization and second level trigger discrimination, the LABRADOR (Large Analog Bandwidth Recorder And Digitizer with Ordered Readout)\cite{labrador} digitization chip, and a motherboard that controls the power, communication and timing. 
A 260-capacitor Switched Capacitor Array (SCA) continuously observes the input RF channels (two channels per antenna) and an additional timing channel. To reduce power consumption and dead time, the information is held and digitized only when a trigger is received. The sampling speed is two giga-samples per second, with a 256 ns buffer depth.
 
The fast and broadband nature of the Askaryan RF signal is exploited for background reduction. Once the voltage measured on an antenna crosses an adjustable threshold, the digitization is triggered and the signal is split into four frequency bands (200-400 MHz, 400-650 MHz, 650-880 MHz and 880-1200 MHz). If enough frequency bands are present in the signal, the channel associated with this antenna will trigger. In the current settings, at least two out of four bands are needed for triggering. The cluster will trigger if enough channels trigger (current setting requires at least three out of four antennas).

An IceCube motherboard controls the timing and communication 
with the surface. A 300 mega-samples per second (3.3 ns per sample) on-board Analog Transient Waveform Digitizer is used for precise trigger timing. 
 To reconstruct an event through the entire array, time resolution of a few 
ns is needed. The RAPCAL (Reciprocal Active Pulsing 
Calibration) method, developed for IceCube is 
used  to achieve that resolution\cite{pdd}.

The antennas used are `fat dipole' antennas, with a bandwidth centered at $\sim400$ MHz in air (or $\sim250$ MHz in ice). Four metal tubes hold the front-end electronics including filters and amplifiers supporting these antennas: specifically, a $450$ MHz notch filter to reject constant noise from the South Pole communication channel, a $200$ MHz high pass filter and a $\sim50$ dB amplifier (see Fig. \ref{fig:frontend}). An additional $\sim20$ dB amplification is done at later stage, for a total of $\sim 70$ dB amplification.  An additional antenna is used as a transmitter for calibration. 
The digitized data is sent to the surface using the IceCube in-ice and surface cables.
Sections 5, 6 and 7 will summarize the results obtained with the 2006-2007 clusters.

\subsection{AURA II: 2008-2009 design}
Three additional radio clusters were built and will be deployed in the coming season (2008-2009). These clusters are based on the above design with a few minor modifications: The new clusters will have two low frequency channels sensitive down to 100 MHz, a stronger ACU and shorter spacing between antennas. Also, the frequency band distribution of the trigger board was modified to three frequency bands (instead of four) and an additional full band.  

The new cluster will allow us to complete the vertexing studies and pinpoint our noise sources, to investigate the South Pole noise environment down to 100 MHz, to perform ice properties measurements, and to look for coincidence events with other South Pole detectors like IceCube, IceTop and RICE. 

\section{Deployment and detector geometry}

\begin{figure}
\begin{center}
\includegraphics*[width=8cm]{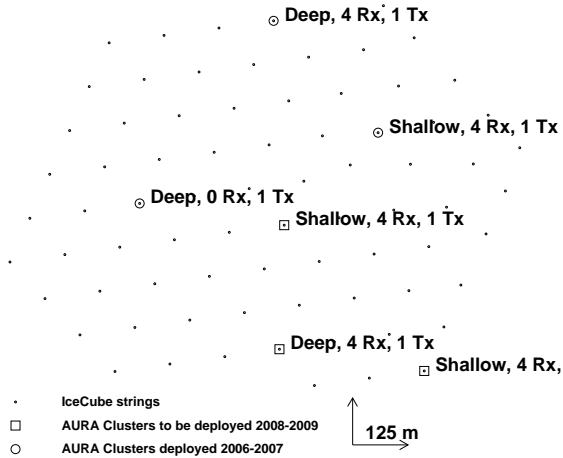}
\end{center}
\caption{Map of the existing and future AURA deployments, plotted on top of the full IceCube array. The distance between adjacent strings is 125 m. ``Deep'' corresponds to $\sim 1400$ m depth, and shallow to $\sim 300$ m depth. The number of receiving channels (Rx) and transmitting ACU channels (Tx) in each clusters are mentioned.  }
\label{fig:aura_geo}
\end{figure}
IceCube's on-going construction activity made it possible to deploy radio clusters down to 1500 m, a depth that is usually less favored by RF detectors due to warmer ice and high drilling cost. In the 2006-2007 season, two AURA clusters were deployed on top of IceCube strings at depths of 1450 m and 250 m. These two clusters will be referred to as ``Deep'' and ``Shallow''.  An additional cluster with a transmitter and no receivers was deployed at 1450 m. 

The clusters have since been taking data. One of the channels of the shallow cluster did not survive the deployment. The waveforms captured on this cluster are strongly saturated. The reason for this is still under investigation, but it might be due to the higher-gain amplifiers used in this cluster and its proximity to the surface. 
In the coming season (2008-2009) three additional clusters will be deployed. 

The locations and depth of the existing and future clusters are shown in Fig. \ref{fig:aura_geo} relative to the IceCube array.The deployment geometry was optimized to cover as large volume as possible and have adequate distances for ice properties studies, constrained by seasonal holes availability set by IceCube's drilling schedule. 

The radio deployment starts after the IceCube string is installed and before the string is lowered to its final depth. For the shallow deployment a special communication cable is used to connect the cluster to the surface. For the deep deployment an extra breakout on the IceCube cable is used. In both cases the IceCube cable is used for mechanical support (see Fig. \ref{fig:cluster-r}). The proximity of the thick cable to the receiving and transmitting antennas causes shadowing of the signal, and the antennas are installed around the support cable, $90^\circ$ apart from each other. Symmetrical antenna design is examined for future use. This antenna will clip on the cable and will have a $2\pi$ sensitivity. 
 
 The data being taken consists of ambient and transient background studies and calibration runs using the AURA transmitter and the in-ice RICE transmitters. Data is being sent to the North daily using a satellite.  

\section{Calibration}
The Array Calibration Unit (ACU) deployed with the radio clusters can emit short pulses (3 ns long) at repetition rates of 20 MHz or less. The signal is expected to be easily detectable over background within $\sim 100$ m from the transmitter. The exact distance depends on the relative polarization of the receiving and transmitting antennas and their location relative to IceCube's surface cable that causes shadowing. The 2006-2007 ACU is not strong enough for inter-cluster studies, and it is used only to study timing and reconstruction within the same cluster. 

\begin{figure}
\begin{center}
\includegraphics*[width=10cm]{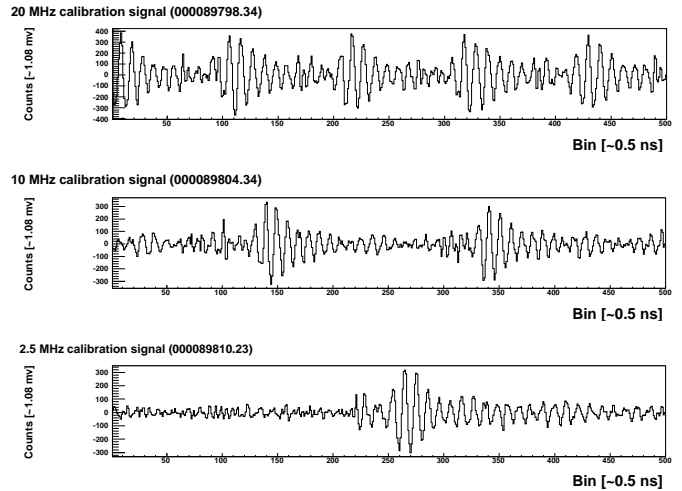}
\end{center}
\caption{Waveforms captured by the deep cluster with different calibration unit repetition rates.The 512 bins cover sampling time of 256ns. Only one channel is shown here.}
\label{fig:acu}
\end{figure}

Figure \ref{fig:acu} shows waveforms taken with the ACU pulsing at different repetition rates, as measured on one of the channels on the same cluster. Due to the finite sampling rate and the filters used, the ACU signal elongates in time up to $\sim40$ ns.  This behavior was verified both in pre-deployment tests in the laboratory and by simulation using the measured gain vs. frequency curve.  

The response of the clusters to fast pulses is extremely important for Askaryan signature searches, since the pulse is  very short ($\sim1$ ns).

In the coming season a stronger ACU unit will be deployed to enable cluster-to-cluster pulsing for ice properties measurement as well as array timing calibration and reconstruction studies. Ultimately a single pulse will trigger the entire array. 


The RICE detector has several buried transmitters capable of producing constant waves (CW) or fast pulses. The CW transmitters were used to verify timing and waveform reconstruction by comparing the measured frequency on AURA to the transmitted frequency. The pulse generator can be used for reconstruction, linearity and ice studies.  Fortunate for some (and unfortunate for others) in order to keep the South Pole environment as quiet as possible, EMI activity is not encouraged; hence the use of strong RF transmitters is restricted, limiting our ability to perform these measurements.

\section{Noise studies}
The proximity of the South Pole station and the IceCube and AMANDA detectors may cause significant RF noise in the frequency band of $200-1200$ MHz where AURA is sensitive. This noise pattern is being carefully studied to evaluate the feasibility of RF experiments at the South Pole, and the ability to distinguish an Askaryan pulse from environmental background.

\begin{figure}
\begin{center}
\noindent
\includegraphics*[width=10cm]{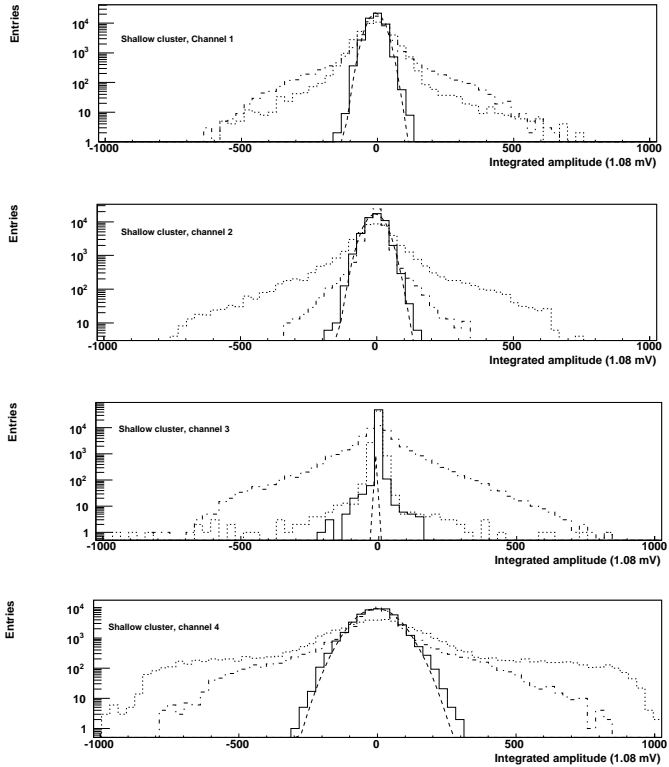}
\end{center}
\caption{Signal distribution for the shallow cluster (overall voltage integrated in waveforms) for ambient background (dashed line) with a fit to a Gaussian (solid). Also shown is the voltage distribution for triggered events in two threshold settings (dotted lines). The third channel of this cluster was damaged during deployment. }
\label{fig:thermal}
\end{figure}

Figure \ref{fig:thermal} shows the voltage distribution for one of the clusters. The ambient background nicely fits a Gaussian with RMS of up to $30$ mV (there are variations between channels). This noise level corresponds to 7 ADC bits depth, leaving 5 bits for signal acquisition. As the threshold for triggering increases, the total power in the triggered waveform increases as expected. The noise spectrum and intensity depends on the location of the antenna relative to the DRM and the type of front-end amplifier used. Background studies were also performed with the IceCube and AMANDA detectors turned off, and though a difference in the trigger rate for the low frequency trigger band was observed, the overall cluster trigger rate was not affected by AMANDA or IceCube.

\begin{figure}
\begin{center}
\includegraphics*[width=10cm]{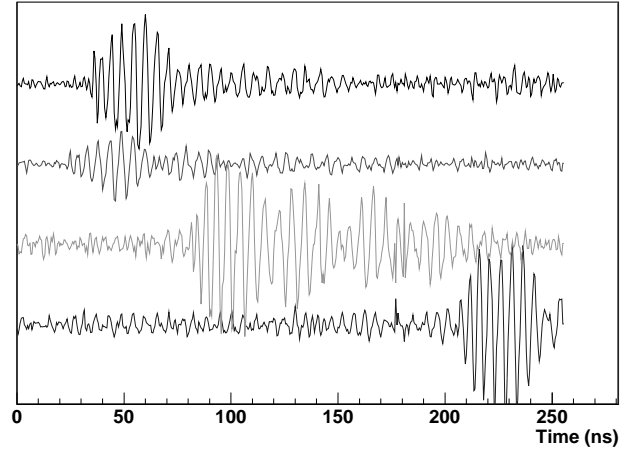}
\end{center}
\caption{A single event for all 4 channels on the deep deployed cluster. The separation between channels is 17 m. The top curve correspond to the signal collected with the top antenna, and the bottom curve to the signal collected on the lowest antenna. The time delays between channels fit a down-going event. Each waveform consist of 512 samples. }
\label{fig:radioWF}
\end{figure}

The fully digitized waveforms allow us to perform reconstruction of the arrival angle of the wave based on the time delays between channels. For example, the event shown in Fig. \ref{fig:radioWF} has a clear down-going signature: since the speed of the signal inside the cable is close to the speed of light in ice, the two top channels record a hit close in time. The two bottom channels record delayed hits.

\begin{figure}
\begin{center}
\noindent
\includegraphics*[width=9cm]{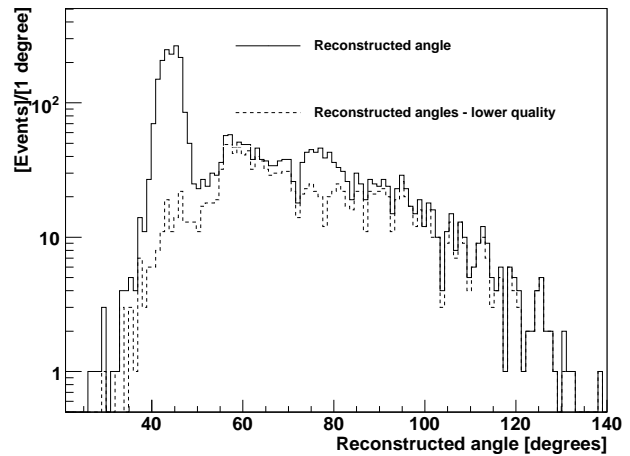}
\end{center}
\caption{Reconstructed angle based on time differences between channels for Deep cluster. The dashed line shows events with lower quality reconstruction. Peaks at $44^\circ$ and $77^\circ$. }
\label{fig:reco_deep}
\end{figure}

\begin{figure}
\begin{center}
\includegraphics*[width=9cm]{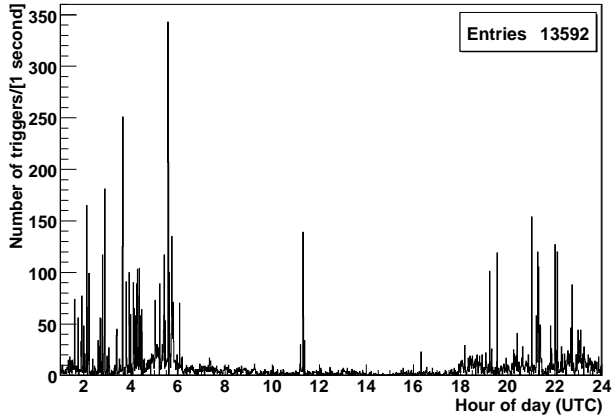}
\end{center}
\caption{Time distribution of triggers at different hours of the day, over 38 days. Higher trigger rates are seen during South Pole active hours, and most of these events trace back to the South Pole station. An exception is a single peak (shown near 11h). This peak is contributed entirely by events taken in a single day, traced back to $77^\circ$. }
\label{fig:radioTime}
\end{figure}

Figure \ref{fig:reco_deep} shows the reconstructed angle for events captured by the deep cluster using a fast first guess algorithm assuming a plane wave. Most of the recorded events seem to be at a zenith angle of $44^\circ$, relative to the deep cluster, tracing back to the South Pole station area. A weaker source at zenith angle of $77^\circ$ of an unknown origin, is also observed. Unlike the noise from the South Pole station, this source generates the measured RF noise only at limited times. Figure \ref{fig:radioTime} shows the number of all triggers at different times of the day over a period of 38 days for the deep cluster. A clear difference is seen between the South Pole station's active and inactive hours. Studies are being made to characterize the time and frequency of these background sources.

The dead channel and the highly saturated signal on the shallow cluster makes the pointing accuracy worse, and nearly impossible for shallow cluster.  
 
\section{Coincident events}
\subsection{AURA clusters coincidence}
For a small subset of down-going events, coincidence between the deep and the shallow clusters was recorded with a time delay of $6.8$ $\mathrm{\mu sec}$ between triggers. This time delay corresponds to a wave propagating directly from the shallow cluster to the deep one. The source can be either the shallow cluster itself or any other noise source sitting next to the line of sight between Deep and Shallow - a good candidate is the South Pole Telescope.

\subsection{Coincidences with other detectors}
\begin{figure}
\begin{center}
\noindent
\includegraphics*[width=10cm]{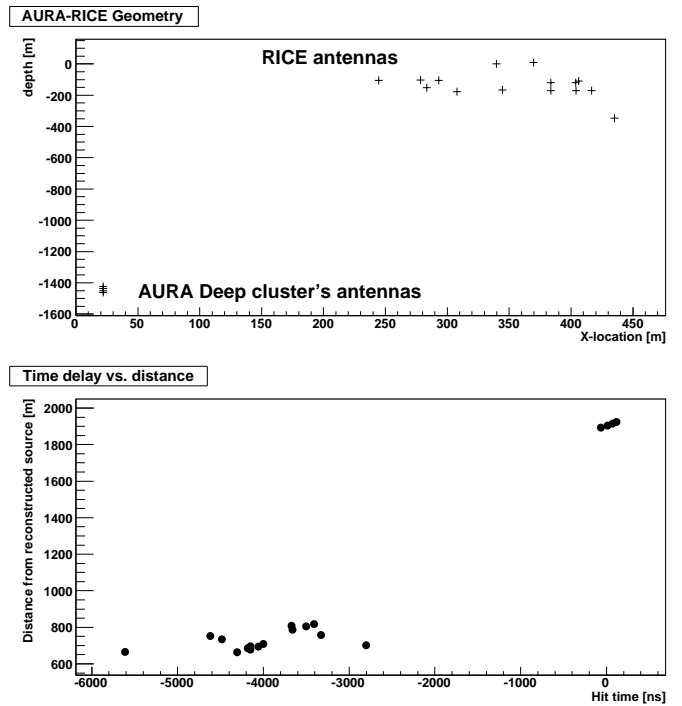}
\end{center}
\caption{The relative geometry (top) of the RICE antennas and the deep AURA cluster (shown x and z only), and the distance between antennas and the reconstructed source position, versus the relative trigger time of the different channels (bottom).}
\label{fig:ricecoinc}
\end{figure}

Some events trigger both the AURA array and the RICE array. This allows 
cross calibration of the detectors and verification of time stamping and 
reconstruction. Figure \ref{fig:ricecoinc} shows the geometry of such an 
event, and the relative time delays. For this event a time delay of $3.498$ $\mathrm{ \mu sec}$ was recorded between the absolute RICE and AURA trigger times. 
The two detectors run separate data acquisition systems. By calculating 
the relative trigger time of each antenna the event can be 
reconstructed. For example, this event was reconstructed back to the 
surface, near to the South Pole station area . All the RICE-AURA events 
are down-going events.

We periodically search for time coincidences between AURA and IceCube or IceTop. 
Though the chances of catching a GZK neutrino inside AURA's volume are 
small, a signal from a lower energy air shower might trigger the 
IceTop array with RF contribution that will propagate into the ice. The 
new low frequency channels should be more sensitive to this.

 \section{Summary and outlook}
The AURA working group of the IceCube collaboration has been studying the possibility of in-ice radio detection of high energy neutrinos. Three clusters were deployed and data are being constantly analyzed to study the noise environment. Coincidences between clusters and with the RICE detector have been shown.

Three additional clusters will be deployed in the coming season with a stronger transmitter allowing us to conduct ice property measurements. With this mini-array we will be able to perform a full point-source reconstruction of noise sources.

A low-cost radio-Cherenkov augmentation to the IceCube detector is being studied to extend IceCube energy sensitivity in the EeV-ZeV range. We would expect to measure a few GZK events per year, some of which are coincident between radio and optical or radio and acoustic
\cite{acoustic} detectors. This detector will consist of several radio sensitive stations deployed at depths between 200-500 m. Studies are underway to design the geometry and optimize the sensitivity of such a detector. See for example \cite{iceray}.  The depth will be chosen based on optimization of  the effective volume covered and the number of holes needed will be based on RF attenuation, ray tracing and drilling costs.   

Due to the low rates of GZK events and the relatively high background, a way to verify that the RF signal originated from a neutrino is to detect hybrid events where a single neutrino also triggered an optical or an acoustic detector \cite{hybrid}. 


\section*{Acknowledgments}
We acknowledge the support from the U.S. National Science Foundation-Office of Polar Program, U.S. National Science Foundation-Physics Division and the University of Wisconsin Alumni Research Foundation.

\end{document}